\documentclass[a4paper,twocolumn,showpacs,amsmath,pra,amssymb,nofootinbib]{revtex4-1} 
\usepackage[T1]{fontenc}
\usepackage[latin9]{inputenc} 
\usepackage{times}
\usepackage{color}
\usepackage{xspace}
\usepackage{amssymb,amsmath}
\usepackage{amsbsy}
\usepackage{graphicx}
\usepackage{epstopdf}
\usepackage{float}
\usepackage{enumerate}

\newcommand{\tS}{t_{\rm{scale}}}
\newcommand{\eqn}[1]{\begin{eqnarray}#1\end{eqnarray}}
\newcommand{\pdiff}[2]{\dfrac{\partial #1}{\partial #2}}
\newcommand{\pdiffl}[2]{{\partial #1}/{\partial #2}}
\newcommand{\half}[1][1]{\frac{#1}{2}\xspace}
\newcommand{\diff}{\mathrm{d}}
\newcommand{\gt}{\ensuremath{g_\mathrm{trap}}\xspace}

\newcommand{\sqrtf}[2]{\sqrt{\frac{#1}{#2}}}
\newcommand{\ob}{\ensuremath{\bar{\omega}}\xspace}
\newcommand{\fracb}[2]{\left(\frac{#1}{#2}\right)}

\newcommand{\bsplit}{\right.\notag\\&\left.}
\newcommand{\od}[1]{\ensuremath{{O\negthickspace\left({{#1}}\right)}}\xspace}
\newcommand{\kt}{\ensuremath{\kb T}\xspace}
\newcommand{\kb}{\ensuremath{k_B}\xspace}
\newcommand{\ve}{\ensuremath{V_\mathrm{eff}}\xspace}
\newcommand{\bose}[3][{}]{\ensuremath{\zeta^{#1}_{#2}\negthinspace\left(#3\right)}\xspace}

\newcommand{\bosee}[3][{}]{\ensuremath{\zeta^{#1}_{#2}\negthinspace\left(e^{#3}\right)}\xspace}
\newcommand{\bx}{\ensuremath{\mathbf{x}}\xspace}

\newcommand{\p}{\mathbf{p}}
\newcommand{\bp}{\mathbf{p}}

\newcommand{\ab}{\bar{a}}
\newcommand{\gsi}{\ensuremath{g_\mathrm{si}}\xspace}
\newcommand{\brac}[1]{\left(#1\right)}
\newcommand{\ght}{g_{\rm{ht}}\xspace}
\newcommand{\muht}{\mu_{\mathrm{ht}}\xspace}
\begin{document} 
\newcommand{\bk}{\mathbf{k}}
\newcommand{\nt}{\tilde{n}}
\newcommand{\Lnh}{\ensuremath{\mathcal{L}}\xspace}
\newcommand{\dbx}{\diff\bx}
\newcommand{\dbk}{\diff\bk}
\newcommand{\nbe}{\ensuremath{\bar{n}_{\mathrm{BE}}}\xspace}
\newcommand{\efht}{\epsilon_{F,\mathrm{ht}}\xspace}
\hyphenation{wave-guide}
\hyphenation{semi-classical}
\title{Geometric scale invariance as a route to macroscopic degeneracy: loading a toroidal trap with a Bose or Fermi gas}
\author{D.~Baillie}  
\author{P.~B.~Blakie} 
\author{A.~S.~Bradley} 

\affiliation{Jack Dodd Centre for Quantum Technology, Department of Physics, University of Otago, Dunedin, New Zealand.}

\begin{abstract} 
An easily scalable toroidal geometry presents an opportunity for creating large scale persistent currents in Bose-Einstein condensates, for studies of the Kibble-Zurek mechanism, and for investigations of toroidally trapped degenerate Fermi gases. We consider in detail the process of isentropic loading of a Bose or Fermi gas from a harmonic trap into the \emph{scale invariant} toroidal regime that exhibits a high degree of system invariance when increasing the radius of the toroid. The heating involved in loading a Bose gas is evaluated analytically and numerically, both above and below the critical temperature. Our numerical calculations treat interactions within the Hartree-Fock-Bogoliubov-Popov theory. Minimal change in degeneracy is observed over a wide range of initial temperatures, and a regime of cooling is identified. The scale invariant property is further investigated analytically by studying the density of states of the system, revealing the robust nature of scale invariance in this trap, for both bosons and fermions. We give analytical results for a Thomas-Fermi treatment. We calculate the heating due to loading a spin-polarized Fermi gas and compare with analytical results for high and low temperature regimes. The Fermi gas is subjected to irreducible heating during loading, caused by the loss of one degree of freedom for thermalization. 
\end{abstract}

 
\maketitle

\section{Introduction}
A range of recent experiments with ultra-cold Bose and Fermi gases  have begun examining finite temperature and non-equilibrium properties in the vicinity of the phase transition, such as critical exponents \cite{Hadzibabic2006a,*Donner07a}, the formation of coherence \cite{Hugbart2007a,*Gerbier04a}, and the production of defects as the system crosses the phase transition \cite{Weiler08a}. As system control is improved, these systems may provide the first quantitative experimental verification of the scaling predicted by the Kibble-Zurek mechanism (KZM)~\cite{Kibble1976,*Zurek1985,*Anglin1999,*Dziarmaga2008a}.

An important and often irreducible aspect of experiments is the inhomogeneous potentials used to confine the gas samples. 
While the most commonly realized potentials are harmonic there has been increasing interest in creating more general shapes, and here we are primarily concerned with toroidal traps. Sagnac interferometry is an important application for large toroidal traps, where the resolution is proportional to the area of the interferometer~\cite{Gustavson00a}. A variety of schemes for making toroidal traps using various combinations of magnetic and optical techniques have been proposed~\cite{Wright00a,*Arnold04a}. Traps of order $\sim 3$ mm diameter have been created~\cite{Gupta2005,*Arnold06a} and used as a waveguide for a Bose-Einstein condensate (BEC). Smaller toroids, in which the gas sample extends completely across the whole potential, have recently been produced~\cite{Weiler08a,*Ryu07a,*Schnelle08a,*Heathcote08a}, and dynamically painted optical potentials~\cite{Henderson09a} offer a new level of flexibility for loading non-harmonic geometries.

 Many features of toroidal geometry have been studied, including topological phases~\cite{Petrosyan99a}, the stability of macroscopic persistent currents~\cite{Javanainen98a,*Benakli99a,*Salasnich99a,*Jackson06a,*Ogren07,*Oegren09} , excitation spectra~\cite{Nugent03a}, atomic phase interference devices~\cite{Anderson03a}, vortex-vortex interactions~\cite{Schulte02a}, generation of excitations via stirring~\cite{Brand01a}, dynamics of sonic horizons~\cite{Jain07a}, parametric amplification of phonons~\cite{Modugno06a}, rotational current generation~\cite{Bhattacherjee04a}, the interplay of interactions and rotation~\cite{Kavoulakis04a}, giant vortices~\cite{Cozzini05a}, and vortex signatures~\cite{Cozzini06a}. Ideal gas theory has recently been used~\cite{Kling08a} to study the rapidly rotating Bose gas in a quartically stabilized harmonic trap realized at ENS~\cite{Bretin04a}. 
 
The ideal gas thermodynamics of a Bose gas in a three dimensional toroidal potential was first considered in \cite{Bradley09a}.  The specific form of toroidal potential examined in that work was the harmonic-Gaussian potential that has been produced using combined magneto-optical traps~\cite{Weiler08a}. A regime was identified in which the thermodynamics of the system were independent of the toroid radius, a feature referred to as \emph{scale invariance}. Scale invariance should allow the efficient loading of a degenerate gas into a large toroidal trap, and provide a system uniquely suitable for quantitatively exploring the critical region.
However, scale invariance could well be highly susceptible to additional effects beyond the ideal gas treatment. In Ref.~\cite{Bradley09a} certain additional effects for the Bose gas were considered: The first order interaction shift to the critical temperature was found to be scale invariant, which motivates a more complete treatment of the interacting system to assess the robustness of the scale invariant predictions, and to critically assess the viability of this system for experimental studies.

\begin{figure}[!t] 
\includegraphics[width=3.4in]{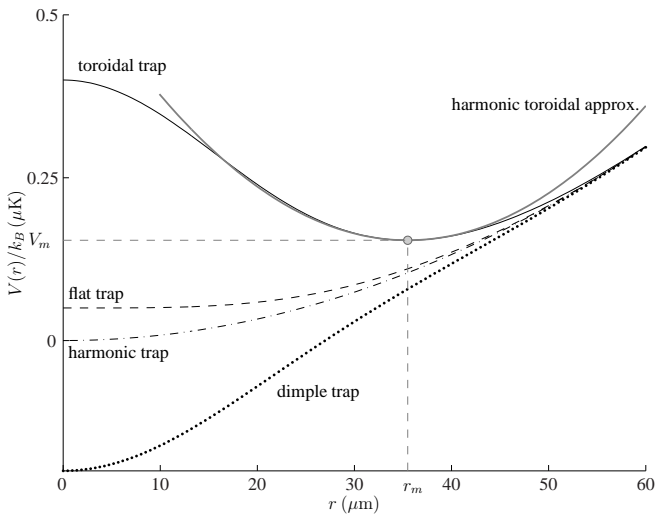}
 \caption{\label{potpic}~Harmonic-Gaussian potential shown at $z=0$ for various values of $V_0$. For $V_0<0$ the potential is an attractive \textit{dimple trap}; for  $0<V_0<V_{\sigma}$ the potential forms a \textit{flat trap} (i.e.~flatter at the origin than the pure harmonic trap); for $V_0>V_\sigma$ a \textit{toroidal trap} is realized with the minimum potential value ($V_m$) occurring at some non-zero radius ($r_m$). The harmonic approximation to the toroidal trap is shown for reference. Results for the case $\sigma_0=24.6\mu$m, $V_{\sigma}/\kb=50$nK, and $V_0/\kb=\{-200$nK$,  50$nK, $400$nK$\}$.}
\end{figure} 
In this work we present a general study of the properties of an ultra-cold gas in the harmonic-Gaussian toroid. We generalize the ideal thermodynamics of \cite{Bradley09a} to a single component fermionic system, and develop a meanfield treatment of the Bose gas to account for interaction effects.  The scale invariant property is further investigated analytically by studying the density of states of the system, revealing the robust nature and fundamental origin of scale invariance in this trap, for both bosons and fermions. We also give an analytical Thomas-Fermi treatment. 
The heating involved in adiabatically loading an ideal gas into a toroidal trap is evaluated analytically in the degenerate and high temperature limits for both the Bose and Fermi systems. 
We compare the analytic predictions for the Bose gas loading against interacting calculations  using self-consistent Hartree-Fock-Bogoliubov-Popov theory. 
Minimal heating is observed over a wide range of initial temperatures, and a regime of cooling is identified. We also compare our analytical predictions for loading a spin polarized Fermi gas into an approximated toroidal trap against a numerical calculation of the full trapping geometry. Heating is found to be unavoidable due to the reduced degrees of freedom available for thermalization in the toroid.

\section{System}
We consider a system of atoms confined in the harmonic-Gaussian potential
\begin{equation}\label{Vdef}
V(\textbf{x})=\frac{m}{2}\left(\omega_r^2r^2+\omega_z^2z^2\right)+V_{0}\exp{\left(-r^2/\sigma_{0}^2\right)},
\end{equation}
where the axial and radial trapping frequencies are $\omega_z$, $\omega_r$, respectively, and   $r$ is the distance from the $z$-axis. This trap can be created using a parabolic magnetic potential, combined with a detuned laser field with Gaussian spatial profile transverse to the propagation ($z$) axis, which forms an optical dipole potential.

Introducing the optical potential is a relatively simple extra step, allowing recent experiments to use this configuration to great effect \cite{Ryu07a,Weiler08a}. This form of harmonic Gaussian trap has an advantage over other realizations of toroids: changes in the toroid size can be easily explored by altering the laser intensity while keeping the harmonic  trap and beam width parameters constant, thus accessing the scale invariant regime~\cite{Bradley09a}.
 
To determine the geometry of the trap it is convenient to define the energy 
\begin{equation}\label{Vs}
V_\sigma\equiv \frac{1}{2}m\omega_r^2\sigma_0^2,
\end{equation}
which is the potential energy of an atom at $r=\sigma_0$ in the harmonic potential. In terms of the radial harmonic oscillator length $a_r=\sqrt{\hbar/m\omega_r}$ it takes the form
\begin{equation}
V_\sigma=\frac{\hbar\omega_r}{2}\left(\frac{\sigma_0}{a_r}\right)^2,
\end{equation}
so that $V_\sigma$ expresses the area of the Gaussian in length and energy units of the radial trap. There are a variety of regimes for the harmonic-Gaussian potential, as summarized in Fig.~\ref{potpic}. Here we will only consider properties of the gas either in the pure harmonic trap ($V_0=0$) or for $V_0 > V_\sigma$, when a true toroidal potential is realized.

The minimum of the toroidal potential has a value of $V_m$ and is located at $z=0$ and $r=r_m$, where
\begin{align}
V_m&=V_\sigma(1+\ln{(V_0/V_\sigma)}),\\
r_m&=\sqrt{\sigma_0^2\ln{(V_0/V_\sigma)}} ,
\end{align}
as shown in Fig.~\ref{potpic}. 

The radial potential can be expanded about $r=r_m$, giving harmonic trapping frequency about the minimum of the toroid
\begin{equation}
\omega_T\equiv\frac{\sqrt{2}\omega_r r_m}{\sigma_0}=\sqrt{2\ln(V_0/V_\sigma)}\omega_r,\label{Eq_omega_T}
\end{equation} which provides the simplest approximate description of the radial degree of freedom. 
The simplest toroidal trap is then given by the \emph{harmonic toroid} 
\begin{equation}
V_{\rm{ht}}(\bx)=\frac{m}{2}\left[\omega_T^2(r-r_m)^2+\omega_z^2z^2\right]+V_m,
\end{equation}
characterized by only three parameters $r_m$, $\omega_z$, and $\omega_T$, which are the radial location of the minimum, and the axial and radial trapping frequencies, respectively (see Fig.~\ref{potpic}). This trap forms a useful approximate description in which analytical progress can be made. We use it extensively in what follows, and compare the results with those calculated for the full potential.

As shown in \cite{Bradley09a}, quantity of fundamental interest for harmonic toroid traps is the geometric mean frequency of the toroid, $\bar{\omega}$, that takes the form
\eqn{\label{wbar}
\bar{\omega}^5&\equiv&\omega_K\omega_z^2\omega_T^2,
}
where $\hbar\omega_K\equiv\hbar^2/2m\pi r_m^2$ is the energy scale associated with azimuthal motion around the toroid~\cite{Bradley09a}. This geometric mean frequency characterizes all ideal gas thermodynamics of the harmonic toroid and is a central physical parameter of the systems studied below.

\section{Bosons and Fermions in a Harmonic toroid}\label{SecHarmToroid}
Scale invariance in the harmonic toroid geometry is a rather specific property that should be introduced with some care. In this section we give a brief review of the result of Ref.~\cite{Bradley09a}: that the harmonic approximation to the harmonic-Gaussian toroidal potential exhibits scale invariance. As a point of comparison we also reiterate some basic properties of the ideal Bose gas in a harmonic toroid. We then provide an analytical Thomas-Fermi treatment for the Bose and Fermi gases. 
 
 \subsection{Scale invariance and density of states}
 As shown in Ref.~\cite{Bradley09a}, the concept of generalized extensive volume introduced by Romero-Rochin~\cite{RomeroRochin05a} can be applied to the harmonic toroid. The extensive volume of the harmonic toroid is
\begin{equation}
{\cal V}=\bar{\omega}^{-5/2},
\end{equation}
and thus a scaling transformation of a particular degree of freedom that preserves ${\cal V}$ can be defined by appropriately changing the other variables to compensate.

Fortunately, the scaling of $\omega_T$ with respect to $V_0$  [see (\ref{Eq_omega_T})] is precisely cancelled by the scaling of $\omega_K$ in Eq.~(\ref{wbar}), giving
\begin{equation}
\bar{\omega}^5=\frac{\hbar}{\pi m \sigma_0^2}\omega_r^2\omega_z^2=\frac{\omega^5}{2\pi V_\sigma/\hbar\omega}.
\end{equation}
The absence of $V_0$ from this expression is the basic scale invariant property of the harmonic toroid: holding all parameters constant while increasing $V_0$ leads to a scaling transformation of the trap geometry which preserves the generalized extensive volume. This has an intuitive explanation: as the toroidal perimeter increases, the radial trapping frequency compensates precisely by increasing so as to preserve ${\cal V}$. 
 
The density of states for the harmonic toroid may be found as~\cite{Bradley09a} 
\begin{equation}\label{htdos}
\ght(\epsilon)= \frac{4\epsilon^{3/2}}{3\sqrt{\pi(\hbar\ob)^5}},
\end{equation}
 where the energy, $\epsilon$, is measured relative to $V_m$.  It is then quite simple to show that scale-invariance holds provided we impose the condition 
 \begin{equation}\label{simpleApprox}
 k_BT\ll V_{\sigma}\ll V_0,
 \end{equation}
which means that the system is in a deep toroid regime described entirely by the density of states (\ref{htdos}).
In this regime the ideal gas properties (e.g.~critical temperature, peak density, entropy) are independent of $V_0$, and hence of the toroidal radius.
  
However, perhaps surprisingly, it was shown in \cite{Bradley09a} that scale-invariance in fact holds in the general harmonic Gaussian trap over a much broader range of parameters, in which the harmonic toroid approximation is not generally applicable. In Appendix~\ref{s:AppSI} we provide a rigorous derivation of the scale invariant property of the full density of states, finding  rigorous criteria to replace the approximate set given in (\ref{simpleApprox}). [See Eq.~\eqref{scaleT}].

\subsection{Bose gas properties}
For any harmonic toroid, a simple expression for the free energy of the boson system can be obtained as~\cite{Bradley09a}
\begin{equation}
{\cal F}=N_0(V_m-\mu)-\frac{\zeta^+_{7/2}(e^{\beta(\mu-V_m)})}{\beta(\beta\hbar\bar{\omega})^{5/2}},
\end{equation}
where $\zeta^{\eta}_\nu(z)=\sum_{k=1}^\infty \eta^{k-1}z^k/k^\nu$ is the polylogarithm function. This expression yields the following estimate for the critical temperature for Bose-Einstein condensation in a harmonic toroid
\begin{equation}\label{TC0}
k_B T_c=\hbar\bar{\omega}\left(\frac{N}{\zeta(5/2)}\right)^{2/5}\simeq 0.89\hbar\bar{\omega}N^{2/5},
\end{equation}
and condensate fraction
\begin{equation}
\frac{N_0}{N}=1-\brac{\frac{T}{T_c}}^{5/2}.
\end{equation}
A more extensive list of thermodynamic quantities of interest is given in Ref.~\cite{Bradley09a}, but we immediately see the consequences of changing geometry through the exponents in these expressions when compared with the harmonic trap, for which $T_c\propto N^{1/3}$, and $N_0/N=1-(T/T_c)^3$.

\subsection{Thomas-Fermi description of the Bose gas}
For a harmonic toroid, where the trap is completely determined by $r_m$, $\omega_z$ and $\omega_T$, and measuring $\mu$ relative to the trap minimum $V_m$, the Thomas-Fermi particle density is 
\eqn{\label{n0}
n_0(\bx)=\frac{\mu}{U_0}\left[1-\left(\frac{r-r_m}{R_T^B}\right)^2-\left(\frac{z}{R_Z^B}\right)^2\right],
}
where this is positive, and zero elsewhere,
where $R_T^B=\sqrt{2\mu/m\omega_T^2}$, and $R_Z^B=\sqrt{2\mu/m\omega_z^2}$ are the Thomas-Fermi radii, and $U_0=4\pi\hbar^2a_s/m$ is the s-wave interaction parameter for scattering length $a_s$. The chemical potential can be expressed in terms of (\ref{wbar}) as
\eqn{\label{mueq}
\muht(N_0)=\hbar\bar{\omega}\left(\sqrt{\frac{8}{\pi}}\frac{a_s}{\bar{a}}N_0\right)^\frac{1}{2},
}
where $\bar{a}=\sqrt{\hbar/m\bar{\omega}}$ is the geometric mean length scale of the toroid. The Thomas-Fermi energy is just $E(N_0)=2\muht(N_0) N_0/3$. Although the density changes with $r_m$, in a system where $\bar{\omega}$ is invariant with $r_m$, the Thomas-Fermi chemical potential and energy are also invariants.
\subsection{Thomas-Fermi description of a zero-temperature spin-polarized Fermi gas}
We consider a spin-polarized single component Fermi gas at $T=0$. Within the standard semi-classical description~\cite{Giorgini08b} a single component gas is described by the non-interacting distribution
\begin{equation}
f(\bx,\p)=\frac{1}{(2\pi\hbar)^3}\frac{1}{e^{\beta(p^2/2m+V(\bx)-\mu)}+1}, 
\end{equation}
and, for example
\begin{equation}
N=\int \frac{d\bx d\p}{(2\pi\hbar)^3}f(\bx,\p)=\int_0^\infty\frac{g(\epsilon)d\epsilon}{e^{\beta(\epsilon-\mu)}+1},
\end{equation}
where $g(\epsilon)$ is the density of states.
For the harmonic toroid density of states (\ref{htdos}), we find the Fermi energy in the harmonic toroid approximation
\begin{equation}\label{ef}
\efht=k_BT_F=\left(\frac{15\sqrt{\pi}}{8}\right)^{2/5}\hbar\bar{\omega}N^{2/5}\simeq 1.62\hbar\bar{\omega}N^{2/5}.
\end{equation}
We see that the scaling with $N$ of the toroidal system is slightly stronger than the $N^{1/3}$ scaling of the harmonic trap \cite{Butts1997a}, and is the same as the scaling of $T_c$ for the ideal Bose gas (\ref{TC0}). The system energy is $E(T=0)=5\efht N/7$. Integrating over momentum, the density distribution is given by
\begin{equation}
n(\bx)=\frac{5}{2\pi}\frac{N}{R_Z^FR_T^FL_T}\left[1-\left(\frac{r-r_m}{R_T^F}\right)^2-\left(\frac{z}{R_Z^F}\right)^2\right]^{3/2},
\end{equation}
where this is positive, and zero elsewhere,
 $L_T=2\pi r_m$ is the toroidal perimeter, $R_T^F=\sqrt{2\efht/m\omega_T^2}$, $R_Z^F=\sqrt{2\efht/m\omega_z^2}$ are the Thomas-Fermi radii. Defining the Fermi momentum $p_F=\sqrt{2m\efht}$, we find the momentum distribution
\begin{equation}
n(\p)=\frac{15}{8\pi}\frac{N}{p_F^3}\left(1-\frac{p^2}{p_F^2}\right).
\end{equation}
While the momentum distribution retains the characteristic spherical shape of the ideal Fermi gas, the power law decay is weaker than the $(1-p^2/p_F^2)^{3/2}$ behavior of the harmonic trap distribution. Since $\efht$ is scale invariant, the peak density in position space, and the momentum distribution are also invariants.

We can also obtain a Sommerfeld expansion of the free energy for the degenerate regime $k_B T\ll \mu$. Evaluating the grand free energy 
\begin{equation}
{\cal F}=k_BT\int_0^\infty d\epsilon\;g(\epsilon)\ln{[1+e^{-\beta(\epsilon-\mu)}]},
\end{equation}
giving, for the harmonic toroid density of states 
\begin{equation}
{\cal F}=-\frac{\zeta^-_{7/2}(e^{\beta(\mu-V_m)})}{\beta(\beta\hbar\bar{\omega})^{5/2}},
\end{equation}
which gives the low temperature expansion, to $\od{\beta\mu}^{-6}$
\begin{equation}\label{Ffermi}
  {\cal F}= -\frac{16}{105}\sqrtf{\mu^7}{\pi(\hbar\bar{\omega})^5}\left[1+\frac{35\pi^2}{24(\beta\mu)^2}+\frac{49\pi^4}{384(\beta\mu)^4}\right],
\end{equation}
which we use below to evaluate isentropic loading.
\section{Bosons and Fermions in a Harmonic-Gaussian trap}
The properties of the ideal Bose gas in the harmonic toroid potential have been studied extensively in \cite{Bradley09a}. In this section we obtain a convenient representation of the grand free energy of the ideal gas in the general harmonic-Gaussian trap that can be readily evaluated for both bosons and fermions. We also find a useful expansion of the density of states that shows the harmonic-Gaussian trap is generally scale invariant, and reveals the necessary conditions for this property.
\subsection{Integral representation of the grand free energy}
The various state variables can be obtained from the  free energy, which for the case of ideal gas  in the harmonic-Gaussian potential is given by 
\begin{equation}\label{FreeEnergy}
  \mathcal{F} 
  =\left\{
  \begin{split}
   N_0(V_m-\mu)-\frac{ \bose[+]{4}{z,\beta V_\sigma, \beta V_0}}{\beta^4 \hbar^3 \omega^3 },&\quad\rm{bosons} \\
    -\frac{ \bose[-]{4}{z,\beta V_\sigma, \beta V_0}}{\beta^4 \hbar^3 \omega^3},&\quad\rm{fermions}
\end{split}
\right.
\end{equation}
where $\beta=1/k_BT$, $z=e^{\beta\mu}$, $\omega^3=\omega_r^2\omega_z$, and
\begin{align}
  \bose[\eta]{\nu}{z,a,b} \equiv a\sum_{k=1}^\infty \frac{\eta^{k-1} z^k}{k^{\nu-1}} \Gamma(ka)\gamma^*(ka,kb),\label{GenBose}
\end{align}
is a generalized $\zeta$-function~\cite{Bradley09a}, 
with $\gamma^*(a,x) \equiv x^{-a}\gamma(a,x)/\Gamma(a)$ and $\gamma(a,x) = \int_0^x e^{-t} t^{a-1} dt$
 the incomplete Gamma function. When $V_0\rightarrow 0$, $ \bose[\eta]{\nu}{z,\beta V_\sigma, \beta V_0}$ reduces to the usual polylogarithm $\bose[\eta]{\nu}{z}$. The expressions for the free energy 
 are obtained by making a semiclassical treatment of the excited levels, and for the Bose case we have explicitly indicated the condensate occupation, $N_0$.  Consistent with the semiclassical approximation we take the condensate mode to have an energy coinciding with the potential minimum, $V_m$. 

As formulated in Eq.~(\ref{GenBose}) the $\bose[\eta]{\nu}{z,a,b}$ is difficult to work with: (i) It is not closed when differentiated with respect to $a$ and $b$ (required for obtaining the entropy and heat capacity from $\mathcal{F}$ etc.), rather leading to an open hierarchy of transcendental functions. (ii) The series in Eq.~(\ref{GenBose}) is formally divergent for fermions when $\mu>V_m$.

However, we have discovered a useful integral representation of $\bose[\eta]{\nu}{z,a,b}$ in terms for the usual Bose and Fermi integrals.  Using
$\gamma(a,x) = x^a  \int_0^1 e^{-xu} u^{a-1}d u$, in Eq.~(\ref{GenBose}), we obtain 
\begin{align}
  \bose[\eta]{\nu}{z,a,b} &= a\int_0^\infty d v\, \bose[\eta]{\nu-1}{ze^{-b e^{-v}-av}}.\label{intbrad}
 \end{align}
Expression (\ref{intbrad}) has the advantage of being closed under differentiation, and is easily evaluated numerically. For example, we have
from $S = -\pdiffl{\mathcal{F}}{T}$
\begin{align}
  \frac{S}{\kb} &= \frac{V_\sigma}{\beta^2(\hbar\omega)^3}\int_0^\infty d v \Big[
  3\bosee[\eta]{3}{\beta(\mu- V_0 e^{-v}-V_\sigma v)}\notag\\
&- \beta \left(\mu-V_0 e^{-v}- V_\sigma v\right) \bosee[\eta]{2}{\beta(\mu- V_0 e^{-v}-V_\sigma v)} \Big].
\end{align}
The representation of Eq.~(\ref{intbrad}) allows all thermodynamic quantities for ideal Bose and Fermi gases to be calculated for all regimes of the harmonic-Gaussian potential (i.e.~flat trap and dimple regimes also), and we use it to calculate the effects of isentropic loading in Section \ref{s:loading}.
\subsection{Generalized scale invariance beyond the harmonic toroid approximation}\label{secgenscalinvproof}
Ideal gas results can be easily determined for the more general harmonic-Gaussian trap once we have an expression for the density of states.
A procedure for systematically extending the density of states to higher order for the harmonic-Gaussian trap is presented in Appendix \ref{AppDOS}. Using this approach, we find the density of states is scale invariant for $\epsilon\le V_0-V_m$ and is given by 
\begin{align}
 g(\epsilon) &=g_{\rm{ht}}(\epsilon)\left[1 
  + \frac{1}{30}\frac{\epsilon}{V_\sigma} + \frac{1}{2520}\fracb{\epsilon}{V_\sigma}^2\right].\label{analyticDOS}
\end{align}
to $\od{{\epsilon}/{V_\sigma}}^3$.  The essence of scale invariance is apparent from this expression: the quantity $V_0$, which sets the radius parameter $r_m$ of the torus, does not appear, despite the distortion of the trap beyond the simple harmonic toroid regime. The swift decay of the coefficients in the series means that for most purposes the first two terms in the expansion are sufficient. 
 \par 
In Appendix \ref{s:AppSI} we use an exact expression for the density of states to show that when $V_0 > V_\sigma$ the density of states is independent of $V_0$ for energies $0\leq \epsilon \le V_0-V_m$, thus scale invariance formally emerges for the full harmonic-Gaussian trap when 
 \begin{align}
 k_BT &\ll V_0-V_m\label{scaleT}.
\end{align}
We note that for a region $ k_BT \ll V_0-V_m$ to exist, $V_\sigma <V_0$ is required and that the approximate condition Eq.~\eqref{simpleApprox} is sufficient, but not necessary for \eqref{scaleT}. Our analysis also shows  that the density of states just above $V_0-V_m$  is almost scale
invariant (the derivative of the density of states with respect to $V_0$ goes to zero
continuously at $\epsilon=V_0-V_m$), and so the temperature restriction (\ref{scaleT}) is not so severe. We thus confirm that the scale invariant behavior of the free energy identified in Ref.~\cite{Bradley09a} arises from the density of states. We also show in Appendix \ref{s:AppSI} that the \emph{trap density of states} is also scale invariant, which shows that interacting extended-Thomas-Fermi and Hartree-Fock-Bogoliubov-Popov (HFBP) equations are scale invariant in the local density approximation, see e.g.~\cite{Baillie10b}.

For the Thomas-Fermi chemical potential, we can see this explicitly by using the first few
terms in $\gt(V)$ given in (\ref{e:gtvpoly}) which gives, for bosons, to $\od{[\hbar\omega/V_\sigma]^{12/5} [a_s/\ab]N_0}^{3/2}$
 \begin{align}
 \mu(N_0) &= \muht(N_0)\Bigg\{
 1 - \frac{1}{18}\left[\frac{1}{2 (2\pi)^{9/10}}\fracb{\hbar\omega}{V_\sigma}^{12/5}\frac{a_s}{\ab}N_0\right]^{1/2}\notag\\&
 + \frac{7}{2592 (2\pi)^{9/10}}\fracb{\hbar\omega}{V_\sigma}^{12/5}   \frac{a_s}{\ab}N_0\Bigg\}. \label{e:muN0V}
 \end{align}
Similarly, for the Fermi energy we find, to $\od{[\hbar\omega/V_\sigma]^{18/5} N^{6/5}}$
\begin{align}
    \epsilon_F &= \efht\left[1-\frac{1}{7}\frac{1}{2^{1/5}60^{3/5}}\fracb{\hbar\omega}{V_\sigma}^{6/5} N^{2/5} \bsplit 
    + \frac{8}{441}\frac{2^{3/5}}{60^{6/5}}\fracb{\hbar\omega}{V_\sigma}^{12/5}N^{4/5} \right].
\end{align}
As these only depend on the invariants $\ab$ and $V_\sigma$, general scale invariance is evident.
\par
We have shown here that scale invariance holds quite generally, both beyond the harmonic toroid approximation, and for degenerate Bose and Fermi systems. Thus it should be possible to load degenerate quantum gases into the scale invariant regime and dilate the toroid to a macroscopic scale. We now assess the level of heating during adiabatic loading from a harmonic trap ($V_0=0$) into tori of different radii.

\section{Isentropic loading\label{s:loading}}
We consider the adiabatic loading of an ultra-cold gas from a harmonic trap into the toroidal trap, and in what follows we denote thermodynamic quantities pertaining to the initial harmonic trap by the subscript $H$ and those pertaining to the harmonic Gaussian toroid by  $T$. 
Under adiabatic loading we equate entropy, i.e.~$S_H(T_H,N)=S_T(T_T,N,V_0)$, with $V_0$ the final height of the Gaussian potential.

\subsection{Bose gas}
\subsubsection{Analytic treatment of the ideal Bose gas}
For the bosonic system, the cases of above and below the critical temperature need to be considered separately, and the relevant degeneracy parameter is the reduced temperature: $t_H=T_H/T_{cH}$,  $t_T=T_T/T_{cT}$, where $T_{cT},T_{cH}$ are the critical temperatures in the initial harmonic trap, and final harmonic toroid trap, respectively.
\paragraph*{Condensed system:} The regime $t_H<1$ and $t_T<1$ was considered in \cite{Bradley09a} where it was shown that 
\begin{equation}
t_T=\left(\frac{8\zeta(4)\zeta(5/2)}{7\zeta(3)\zeta(7/2)}\right)^{2/5}t_H^{6/5} \approx 1.08 t_H^{6/5}.\label{eq:TH}
\end{equation}

\paragraph*{Uncondensed system:} For the regime $t_H>1$ and $t_T>1$ we can make a Boltzmann approximation.
In this regime we have  that $S_H/N\kb \rightarrow 4+\ln \zeta(3) + 3 \ln t_H$ and (for the harmonic toroid) $S_T/N\kb \rightarrow \frac{7}{2} - \ln \zeta(5/2) + \frac{5}{2}\ln t_T$, giving
\begin{align}\label{eq:aboveTc}
  t_T = \left(\frac{\zeta(5/2)}{\zeta(3)}\right)^{2/5}\hspace{-1mm}e^{1/5} t_H^{6/5} \approx 1.28 t_H^{6/5},
\end{align}
with the same power as in the condensed system case. While this analysis breaks down very close to the critical temperature (where the harmonic system may be condensed, yet the toroidal system uncondensed), it provides a useful upper bound for the jump in relative temperature that can occur in the vicinity of the critical point. Taking the difference between (\ref{eq:TH}) and (\ref{eq:aboveTc}) for $t_H=1$, we find $\Delta t_T(t_H=1)= 0.19$, i.e. a maximum change of relative temperature of less than 20\%.

\subsubsection{Meanfield treatment of the interacting Bose gas}
We model interactions using a HFBP treatment, which is a mean-field approximation, valid in the weakly-interacting regime. HFBP allows for interactions between the condensate and non-condensate and `gives a good approximation at all temperatures' \cite{Griffin96}. 

We make a local density approximation to the HFBP, which is valid when the temperature is large compared to the energy spacing. Our approach is similar to the harmonic oscillator work of Giorgini et al., \cite{Giorgini97b}, except that we use the generalized Thomas-Fermi approximation for the condensate density
\begin{align}
    n_0(\bx) &= \frac{1}{U_0}\left[\mu-V(\bx)-2U_0\nt(\bx)\right],\label{e:n0x}
\end{align}
where this is positive, and zero elsewhere,
and we include the quantum depletion, giving
\begin{align}
\nt(\bx) 
  &=\int  \frac{d\bp}{(2\pi\hbar)^3} \left\{  \frac{\Lnh(\bx,\bp)}{E(\bx,\bp)}\nbe[E(\bx,\bp)] \bsplit
  + \frac{\Lnh(\bx,\bp) - E(\bx,\bp)}{2E(\bx,\bp)} \right\},\label{e:nbdintk}
\end{align}
where  $\nbe(E)=(e^{\beta E} -1)^{-1}$  and $\Lnh(\bx,\bp)$ and $E(\bx,\bp)$ are the Hartree-Fock and HFBP energy dispersions in the local density approximation
\begin{align}
    \Lnh(\bx,\bp) &= \frac{p^2}{2m} + V(\bx) -\mu + 2U_0[n_0(\bx)+ \nt(\bx)],\\
E(\bx,\bp) &= \sqrt{\Lnh^2(\bx,\bp)-\left[U_0 n_0(\bx)\right]^2}.
\end{align}
We calculate the total condensate and non-condensate numbers as
\begin{align}
    N_0 = \int d\bx \, n_0(\bx),\hspace{5mm} \tilde{N}= \int d\bx \, \nt(\bx)\label{e:Nt}.
\end{align}

\begin{figure}[!t] 
\includegraphics[width=3.4in]{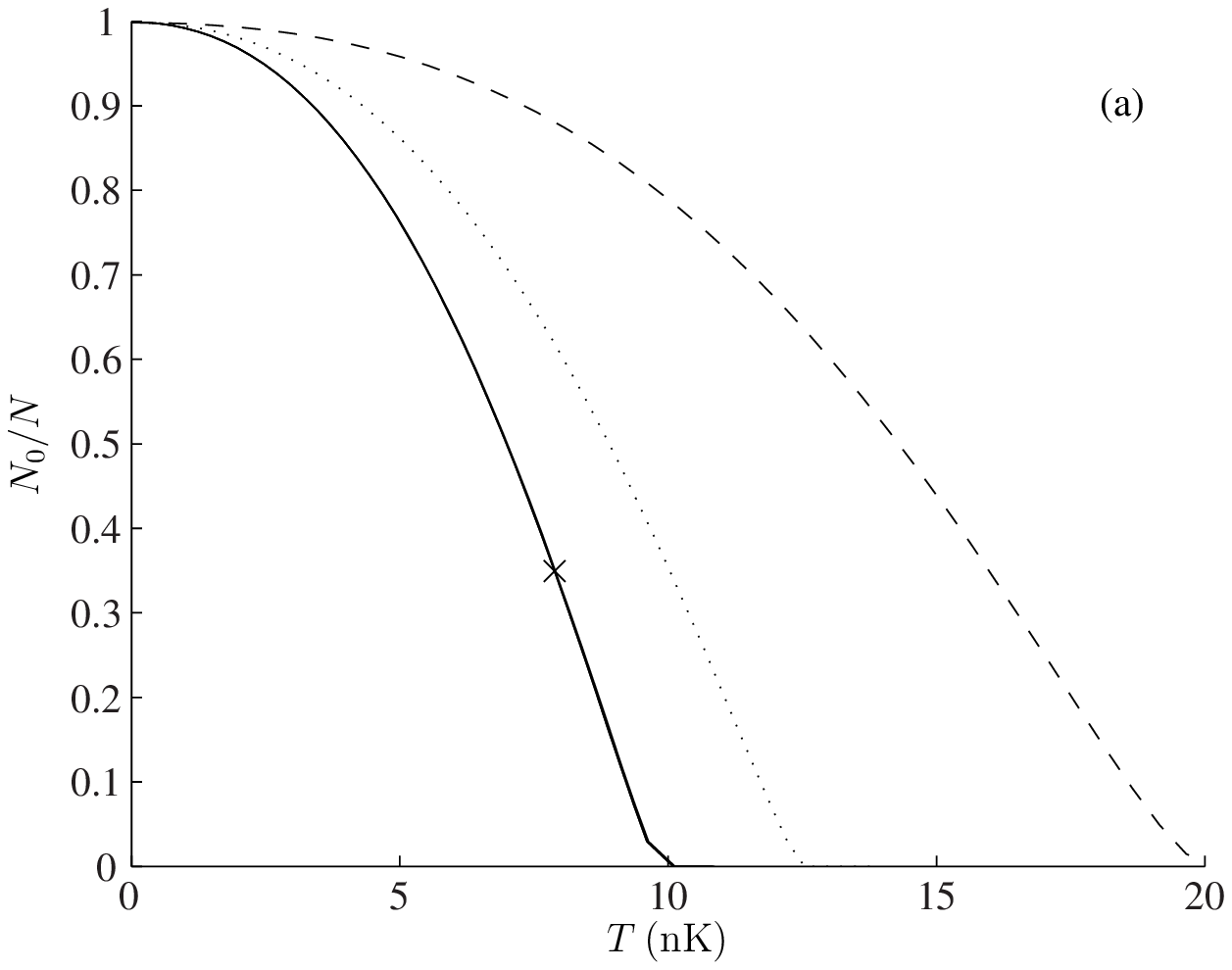}
\includegraphics[width=3.4in]{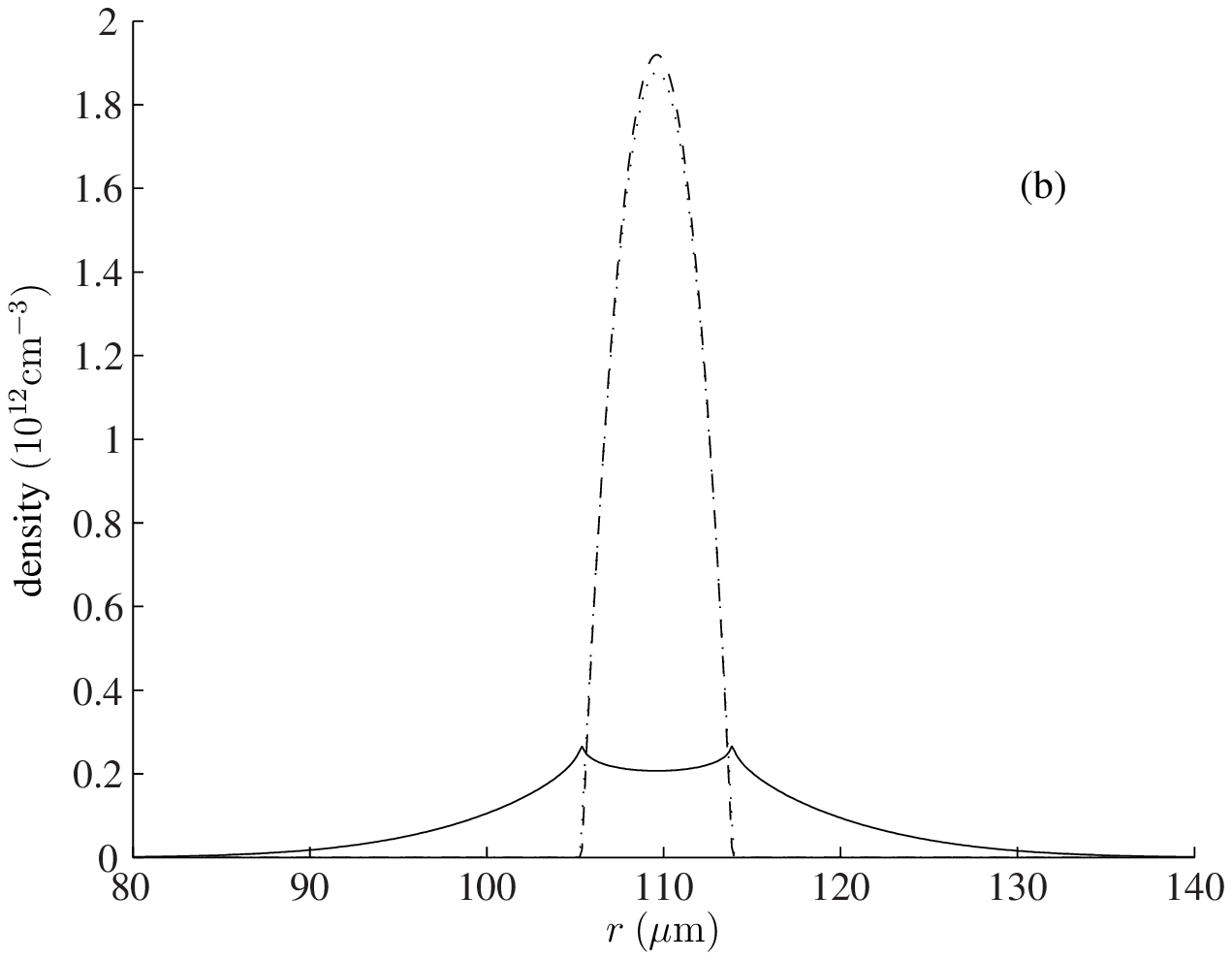} 
\caption{\label{f:HFBsoln}(a) Condensate fraction versus temperature for fixed
$V_\sigma/\kb = 94 \mathrm{nK}$ and harmonic trap (dashed), $V_0/V_\sigma = 1$ (dotted), 5, 10 
 and harmonic toroid (all solid and indistinguishable on this scale). The $\times$  indicates the results studied in subplot~(b). 
 The critical temperatures are 20, 13 and 10 nK respectively.
(b) Radial density at $z=0$ for the non-condensate (solid), condensate (dashed) and pure Thomas Fermi (dotted) [from Eq. \eqref{n0} with $\mu(N_0)$ from \eqref{mueq} evaluated for the same $N_0$ as the finite temperature system]. Density profiles are shown for the case indicated with $\times$ in (a), at $T = 0.8T_c$ with 
$V_\sigma/\kb = 94 \mathrm{nK}$ and  $V_0/V_\sigma=5$. Calculations are for a system of $10^5$ $^{87}$Rb atoms ($a_s=5.77\:\mathrm{nm}$), with $[\omega_z, \omega_r] = 2\pi [15.3, 7.8]\:\mathrm{Hz}$ (so $V_\sigma = 200\hbar\omega$).}
\end{figure}

We solve Eqs.~\eqref{e:n0x}--\eqref{e:Nt} self-consistently, finding $\mu$ so that $N=N_0 + \tilde{N}$, for fixed $N, m, \omega_r, \omega_z, V_0, \sigma_0$ and $a_s$. Examples of our HFBP results are shown in Fig.~\ref{f:HFBsoln}. In Fig.~\ref{f:HFBsoln}(a) we see the effect of increasing $V_0/V_\sigma$ on the condensate fraction versus temperature. Notably we see a suppression of the critical temperature $T_{c}$ with increasing $V_0$, and the emergence of scale invariance, whereby the results for $V_0/V_\sigma=5 $ and $10$ are visually indistinguishable. In Fig.~\ref{f:HFBsoln}(b) we show a typical example of the condensate and non-condensate density in the toroidal  potential.

An important effect of interactions is the depletion of the condensate at zero temperature, known as the {quantum depletion}. While the quantum depletion is generally very small in harmonic traps (typically $<1\%$ \cite{Dalfovo1999}),  other potentials such as lattices can considerably enhance this \cite{Xu2006a,Baillie2009b}. By visual inspection of  Fig.~\ref{f:HFBsoln}(a) it is clear that the quantum depletion is not significantly enhanced in the toroid. We can gain some approximate quantitative understanding of this as follows: Using $\nt(\bx)= (mU_0n_c(\bx))^{3/2}/3\pi^2\hbar^3$ for the depletion at $T=0$
from \cite{Pitaevskii99}, to lowest order $U_0n_c(\bx)\approx \muht(N) -
V(\bx)$, using the chemical potential from \eqref{mueq}, and the harmonic-toroid density
of states from the first term of \eqref{e:gtvpoly}, we get in the harmonic toroid limit
\begin{equation}\tilde{N}_{T=0} =  \frac{2^{35/8}}{15\pi^{9/8}} \left(\frac{a_s}{\bar{a}}N\right)^{5/4} \approx 0.382\left(\frac{a_s}{\bar{a}}N\right)^{5/4},\label{e:htdepletion}\end{equation}
compared to $\tilde{N}_{T=0} = 3^{1/5}5^{6/5}(Na_s/a)^{6/5}/2^{9/2} \approx 0.380 (Na_s/a)^{6/5}$ for the harmonic
trap, where $a\equiv\sqrt{\hbar/m\omega}$ is the geometric mean harmonic oscillator length. We find that these estimates agree well with the full HFBP calculations as shown in Table~\ref{t:depletion}.

\begin{table}[!t]
\begin{center} {\footnotesize
\begin{tabular}{l|ccccc}
\hline
\hfill$V_0/V_\sigma$  & 0 & 1 & 5 & 10 &  harmonic toroid \\ \hline\hline
Numerical HFBP & 174 & 110 & 93 & 93 & 93 \\
Harmonic approximations& 177 & & & & 94 \\ \hline
\end{tabular} }
\end{center}
\caption{\label{t:depletion}  Numerical HFBP predictions of depletion at zero temperature for the cases in Fig. \ref{f:HFBsoln}(a). The analytical estimates for a purely harmonic trap and a harmonic toroid are given in the bottom row.}
\end{table}

\subsubsection{Application of meanfield treatment to isentropic loading of an interacting Bose gas}
Using our meanfield theory we consider the effect of isentropic loading on the system degeneracy. To do this we initially determine the entropy of a harmonically trapped sample at temperature $T_H$, using  
\begin{align}
  S &=\hspace{-1mm} \int\hspace{-1mm}\frac{d\bx\,d\bp}{(2\pi\hbar)^3}\hspace{-1mm} \left\{\beta E(\bx,\bp)\nbe[E(\bx,\bp)] 
    \hspace{-1mm} -\hspace{-1mm}\ln\left[1-e^{-\beta E(\bx,\bp)}\right]\right\}.\label{entropy}
\end{align}
Then, for each toroidal configuration of interest we find the temperature ($T_T$) at which the same entropy is obtained [again using Eq.~(\ref{entropy})]. This determines the isentropic mapping $T_H\to T_T$ that characterizes how the absolute temperature changes during loading. However, we are more interested in the change in degeneracy that occurs,  i.e., the change in reduced temperature, and for this purpose we use the critical temperatures $T_{cH}$ and $T_{cT}$ self-consistently determined by the HFBP calculation. 

\begin{figure}[!t]
\includegraphics[width=3.4in]{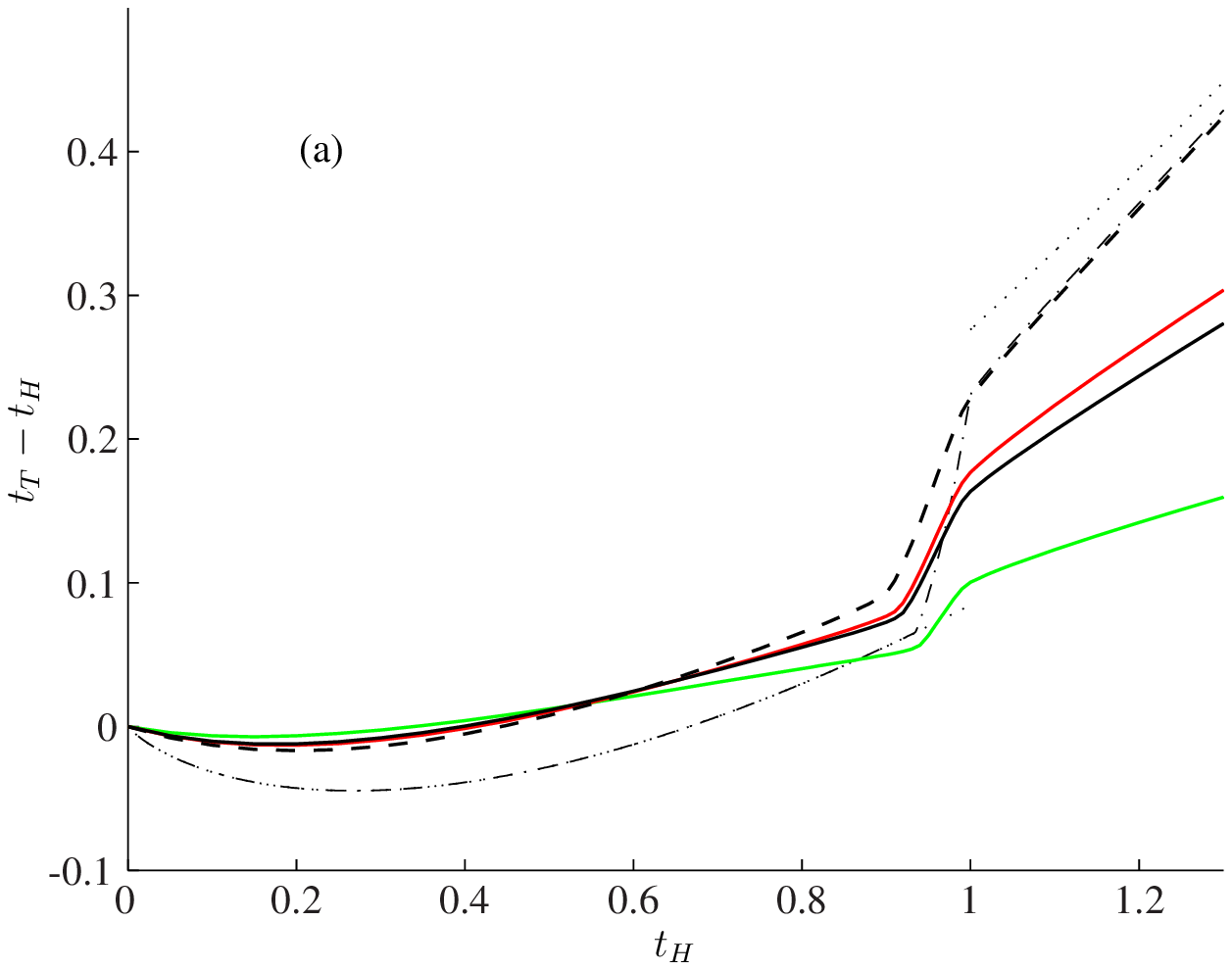}
\includegraphics[width=3.4in]{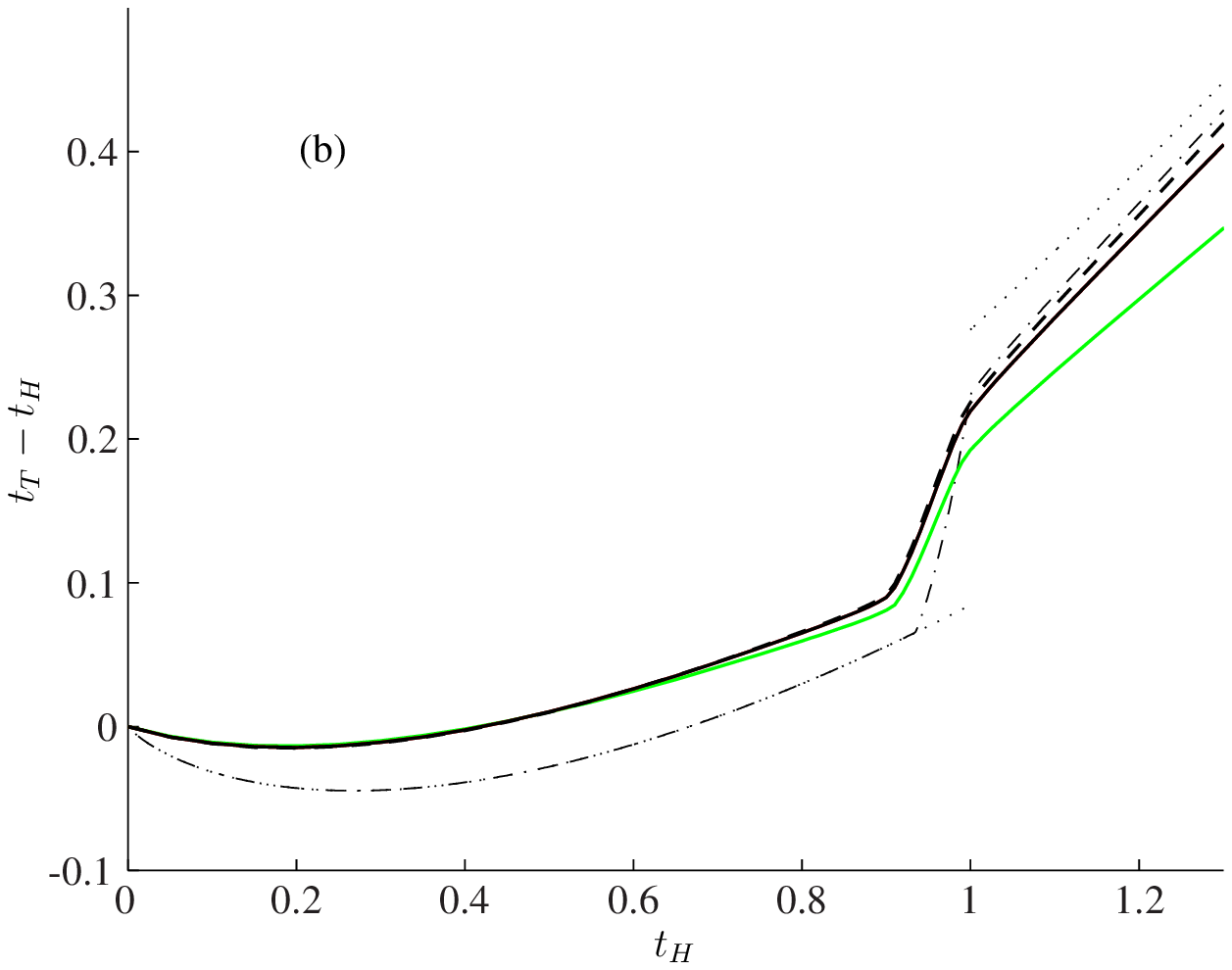}
\caption{\label{f:loading}(Color online) Effect on the reduced temperature of adiabatically loading bosons into a toroid with (a) $V_{\sigma}=20\hbar\omega$, (b) $V_{\sigma}=200\hbar\omega$. Other parameters are for $N=10^5$ $^{87}{\rm Rb}$ atoms with 
$[\omega_z, \omega_r] = 2\pi [15.3, 7.8]\:\mathrm{Hz}$, and $V_0/V_\sigma = 1$ (green/light-grey curve), 5 (red/dark-grey curve), 10 (black curve), and the harmonic toroid approximation (dashed curve). Also shown are the exact non-interacting results for the harmonic Gaussian trap (dashed-dotted curve), the harmonic toroid result Eq.~\eqref{eq:TH} (dotted curve for $t_H<1$), and the limits Eq.~\eqref{eq:aboveTc} (dotted curve for $t_H>1$). For (b), $V_0/V_\sigma=5$ and $V_0/V_\sigma=10$ are indistinguishable on this scale.}
\end{figure}

Our results for the change in degeneracy are shown in Fig.~\ref{f:loading} for the case of a system of $10^5$ $^{87}$Rb atoms in two different toroidal configurations. Here we compare the 
change in reduced temperature that occurs as a function of the initial reduced temperature in the harmonic trap.

A characteristic feature of all curves is the distinctive steep middle segment that occurs near $t_H\sim 1$. This arises as the gas passes through the critical point, noting that the critical point entropy is lower in the toroid than the harmonic trap (see Tab.~\ref{numericalparameters}). Thus as the initial temperature $t_H$ (and hence entropy) increases, a point is reached (the beginning of the steep region) where isentropic loading will transform a partially condensed system in the harmonic trap into a normal system in the toroid. The steep region then terminates when $t_H=1$ and the harmonically trapped gas also becomes normal.

\begin{table}[!t]
\begin{center} {\footnotesize
\begin{tabular}{l|ccccc}
\hline
$\qquad\qquad V_0/V_\sigma$  & 0 & 1 & 5 & 10 & harmonic toroid\\ \hline\hline
(a) $V_{\sigma}=20\hbar\omega$&     &&&&   \\  
$k_BT_c/V_{\sigma}$ &    2.2 & 1.9 & 1.6 & 1.6 & 1.7 \\ 
$S(T_c)/Nk_B$        &    3.6 & 3.2 & 3.0 & 3.0 & 2.9 \\ 
$k_BT_F/V_{\sigma}$ &    4.2 & 3.6 & 3.0 & 3.0 & 3.1 \\ 
$S(T_F)/Nk_B$        &    5.8 & 5.4 & 5.0 & 5.0 & 4.7 \\ \hline
(b) $V_{\sigma}=200\hbar\omega$&     &&&&   \\  
$k_BT_c/V_{\sigma}$ &    0.22 & 0.13 & 0.11 & 0.11 & 0.11 \\ 
$S(T_c)/Nk_B$        &    3.6 & 3.0 & 2.9 & 2.9 & 2.9 \\ 
$k_BT_F/V_{\sigma}$ &    0.42 & 0.25 & 0.19 & 0.19 & 0.19 \\ 
$S(T_F)/Nk_B$        &    5.8 & 4.9 & 4.8 & 4.8 & 4.7 \\ \hline
 \end{tabular} }
\end{center}
\caption{\label{numericalparameters} 
Numerically obtained entropy and degeneracy temperatures for the systems studied in Sec.~\ref{s:loading}.}
\end{table}

The ideal harmonic toroid result (\ref{eq:TH}) predicts $t_H^*=0.67$ as the temperature for which the degeneracy is unchanged upon loading (i.e.~$t_T=t_H^*$). For  $t_H<t_H^*$ Eq.~(\ref{eq:TH}) predicts that loading cools the system, and for  $t_H>t_H^*$ heating will occur (see dotted curve in Fig.~\ref{f:loading}). In general we can identify $t_H^*$ as being where the curves in Fig.~\ref{f:loading} intercept the horizontal axis (i.e.~where $t_T-t_H=0$). So, by inspection of our results, we see that interaction effects strongly suppress the amount of cooling that occurs and reduces the value of $t_H^*$ significantly (e.g.~$t_H^*\approx 0.4$  in the harmonic-toroid limit).
We note that for $t_H\lesssim1$ the change in relative temperature that occurs during loading is generally quite small. Thus the initial harmonic trap (in which temperature is easy to measure) is a good thermometer of the toroid in this degenerate regime and should provide a convenient mechanism for accurately preparing a loaded toroid at target degeneracy.

For cases where the toroid is above its critical temperature the effect of interactions is much less pronounced, in the sense that the interacting result for the harmonic toroid is in good agreement with the noninteracting result in the same limit.
However, comparing the results between Fig.~\ref{f:loading}(a)-(b) we see that the rate of convergence to the scale invariant result (i.e.~as $V_0/V_{\sigma}\to\infty$) varies appreciably. For the parameters in Fig.~\ref{f:loading}(a)  the critical temperature is quite high (see Tab.~\ref{numericalparameters}) and scale invariance emerges only when $V_0$ is significantly larger than this [see Eq.~(\ref{scaleT})]. For the parameters in Fig.~\ref{f:loading}(b) the critical temperature is lower (see Tab.~\ref{numericalparameters}) and the system is seen to approach  scale invariance  behavior more rapidly with increasing $V_0$. 

\subsection{Fermi gas}
\subsubsection{Analytic treatment of the ideal Fermi gas for the harmonic toroid}
\paragraph*{Degenerate system:} The entropy is given by $S=-\partial_T{\cal F}|_{\mu,\omega}$. Using Eq.~\eqref{Ffermi}, in the low temperature limit we find to $\od{\beta\mu}^{-4}$
\begin{equation}
  \frac{S_T}{\kb}=\frac{4}{9\beta}\sqrtf{(\pi\mu)^{3}}{(\hbar\bar{\omega})^5}\left[1+\frac{7\pi^2}{40(\beta\mu)^2}\right].
\end{equation}
Written in terms of the Fermi temperature in the toroid, $T_{FT}$ [see Eq.~(\ref{ef})], to lowest order in $T_T$, we find for the harmonic toroid
\begin{equation}\label{Sft}
  \frac{S_T}{N k_B}\simeq \frac{5\pi^2}{6}\frac{T_T}{T_{FT}}.
\end{equation}
The expression for the harmonic trap~\cite{Carr04a} reads
$S_H/Nk_B\simeq \pi^2{T_H}/{T_{FH}}$,
where $T_{FH}=(6N)^{1/3}\hbar\omega/k_B$ is the Fermi temperature. In terms of the reduced temperatures $t_H=T_H/T_{FH}$, and $t_T=T_T/T_{FT}$, we find that isentropic loading gives the reduced temperature 
\begin{equation}\label{lowTlim}
t_T=\frac{6t_H}{5},
\end{equation} 
reflecting the reduction in available degrees of freedom in the toroidal system. Thus, in contrast to the Bose gas, there does not exist a regime in which the Fermi gas undergoes cooling through loading into a harmonic toroid.
\paragraph*{High temperature system:}
Using the Boltzmann approximation, we find the entropy in the high temperature regime $S_T/N\kb = 7/2+\ln(15\sqrt{\pi}/8) + \frac{5}{2}\ln t_T$ which, with harmonic trap expression, $S_H/N\kb = 4+\ln 6 + 3\ln t_H$, gives the relation
\begin{equation}
    t_T=\fracb{256 e}{25\pi}^{1/5} t_H^{6/5} \simeq 1.55 t_H^{6/5}\label{e:boltzT},
\end{equation}
which has the same exponent as found for Bose gas loading.
\begin{figure}[!t]
\includegraphics[width=3.4in]{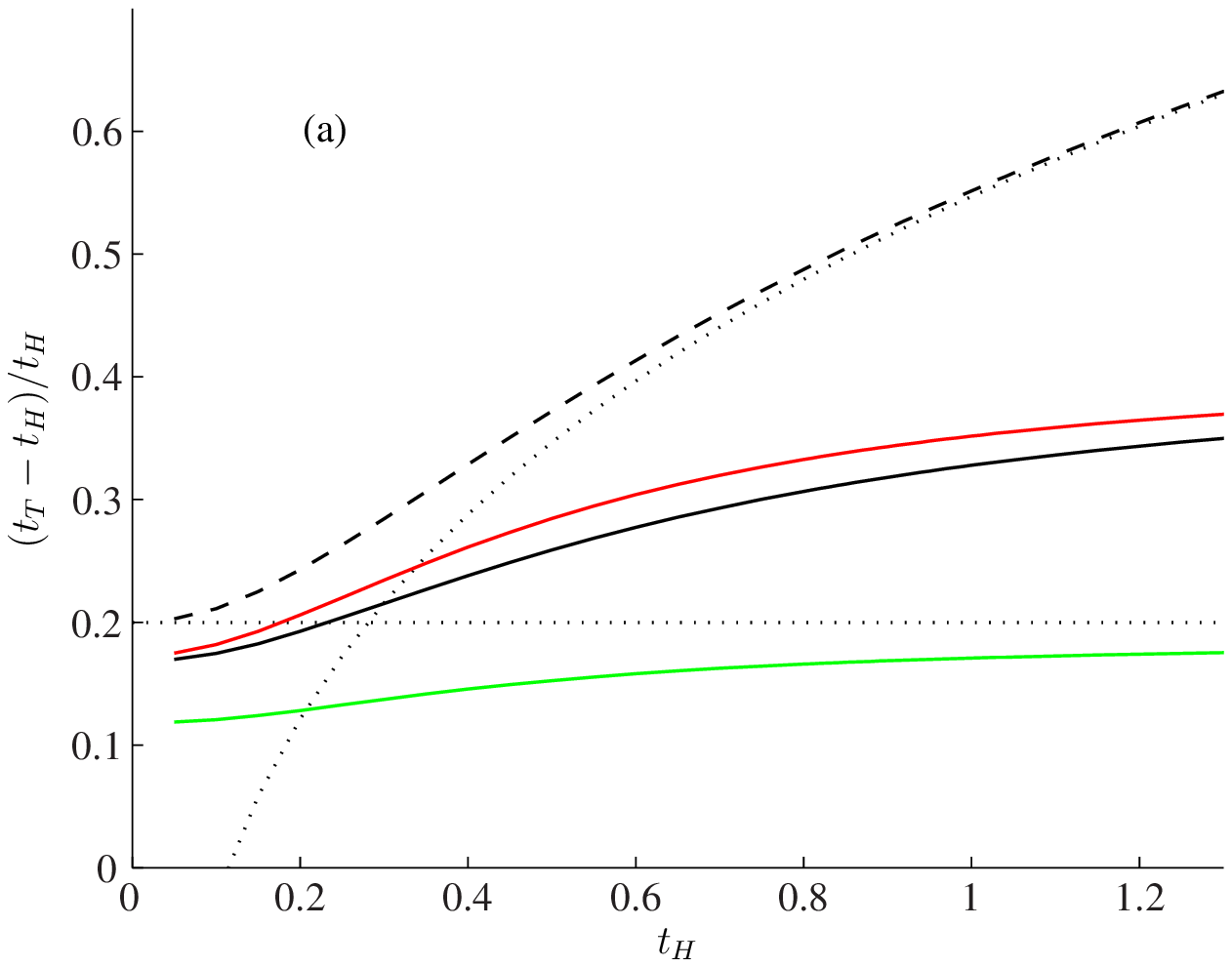}
\includegraphics[width=3.4in]{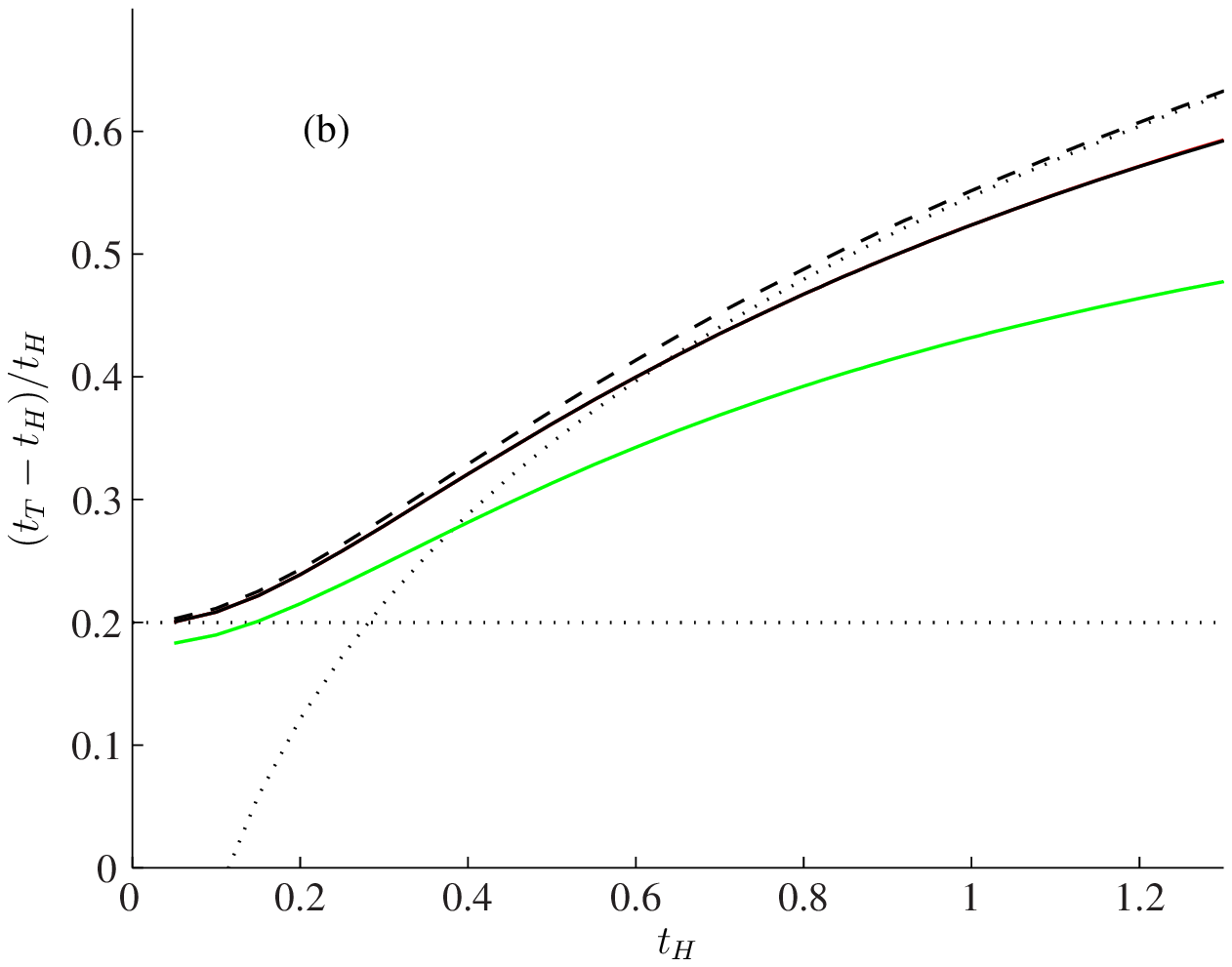}
\caption{\label{f:loadingF}(Color online) Relative effect on the reduced temperature of adiabatic loading a spin polarized Fermi gas into a harmonic-Gaussian toroid for (a) $V_\sigma = 20\hbar\omega$ and (b) $V_\sigma = 200\hbar\omega$. Numerical results are shown for $V_0/V_\sigma = 1$ (green/light-grey curve), 5 (red/dark-grey curve), 10 (black curve) and the harmonic-toroid limit (dashed curve). Also shown are the Sommerfeld (horizontal dotted curve, from \eqref{lowTlim}) and Boltzmann (dotted curve, from \eqref{e:boltzT}) approximations for the harmonic toroid.  For (b) $V_0/V_\sigma=5$ and $10$ are almost indistinguishable on this scale.}
\end{figure}

\subsubsection{Comparison with numerical treatment of the ideal Fermi gas}
Having established the high and low temperature behavior of a spin polarized Fermi gas under isentropic loading into a harmonic toroid, in this section we numerically determine the heating caused by loading into the general harmonic 
Gaussian toroid using  the free energy (\ref{FreeEnergy}), making use of the representation (\ref{intbrad}). In the appropriate limits we verify the analytical results given above. 

For the Fermi gas it is convenient to consider the relative change in reduced temperature that occurs under isentropic loading. Results shown in Fig.~\ref{f:loadingF}, are for the same atom number ($N=10^5$) and trap configurations as the bosonic results in Fig.~\ref{f:loading}.
We note that, in  Fig.~\ref{f:loadingF}(a) the zero temperature limit is markedly different from the harmonic toroid result. Including the next term in the scale invariant density of states, \eqref{analyticDOS}, gives the following expansion of \eqref{lowTlim}, to $\od{N/(V_\sigma/\hbar\omega)^3}^{4/5}$
\begin{align}
  \frac{t_T}{t_H}  &= \frac{6}{5} - \frac{1}{7}\fracb{3N}{1250(V_\sigma/\hbar\omega)^3}^{2/5}.
\end{align}
We see that, while harmonic toroid description is not particularly good in Fig.~\ref{f:loadingF}(a) due to the smallness of $V_\sigma$, the reduced temperature still approaches a scale invariant limit at low temperature. The heating incurred when loading into a system with small $V_\sigma$ is \emph{less} than that seen for loading into a harmonic toroid. 

The rate of convergence to the scale invariant result with increasing $V_0/V_\sigma$ is slower in Fig.~\ref{f:loadingF}(a) than for bosons in Fig.~\ref{f:loading}(a). Part of this effect is only apparent, due to Fig.~\ref{f:loadingF} showing the \emph{relative} change in reduced temperature. The actual effect is due to the fact that, while the bosonic critical temperature (\ref{TC0}) and the Fermi temperature (\ref{ef}) have the same dependence on physical parameters, the prefactor for the Fermi energy is almost twice as large (see Table~\ref{numericalparameters}), and also due to the Fermi distribution: the Fermi temperature itself is an important energy scale for the system, and unless
\begin{equation}k_BT_{FT}\lesssim V_0-V_m,\label{FermiTscaleinvariance}
\end{equation} 
the system will be have occupied levels extending to regions of the spectrum that are not scale invariant. 
\section{Discussion and conclusions}
\subsection{Limitations of adiabaticity}
Any defect in the trapping geometry may have a significant effect when interacting with, for example, a persistent current. The scale invariant properties we have emphasized here rely crucially on having a very clean trapping geometry for any scaling transformations of the toroid to preserve the state of the system. While optical traps created with continuous laser light can be difficult to refine to a perfect single mode Gaussian state, recent work on alternative procedures to produce lattice potentials has demonstrated a way to make very smooth (defect free) optical potentials with pulsed femtosecond lasers~\cite{Bakr09a}.

If the system is in the scale invariant regime, conditions for adiabaticity when scaling the toroid may be much less stringent. The adiabaticity criterion for scaling in time $\tS$ would normally be $2\pi/\omega_K\ll \tS$, which, for large tori, would impose an exceedingly long timescale, as $\omega_K\sim \hbar/mL^2$, for toroidal circumference $L$. For $^{87}$Rb, the scaling time for $L\sim 1{\rm mm}$ is $\tS\gtrsim 10^3$s. 
However, to the extent that the potential can be engineered to be highly cylindrically symmetric, a radial scaling will only couple to radial excitations of the gas. Thus, the criteria relaxes to $2\pi/\omega_T\ll\tS$ once the system is in the harmonic toroid regime, and it will usually be the case for a large toroid that $\omega_T^{-1}\ll \omega_K^{-1}$. For example $\omega_K/\omega_T \sim 10^{-4}$ for the parameters of Fig. \ref{f:HFBsoln}(b).
\subsection{Limitations on toroid size}
As the toroid size is increased, the largest energy scale in the system becomes $\hbar\omega_T$. In order that many radial trap states are occupied (where the theory we have developed is applicable), the appropriate energy scale of the system must be large compared to $\hbar\omega_T$:

\begin{enumerate}[(i)]
 \item For bosons at zero temperature, using \eqref{mueq} to set $\muht(N) \gg \hbar\omega_T$, we get
\begin{align}
    r_m \le \frac{2^{1/4}}{\sqrt{\pi}} \sqrt{a_s\sigma_0}\fracb{\omega}{\omega_r}^{3/2}\hspace{-3mm}  N^{1/2}
    \approx .671 \sqrt{a_s\sigma_0}\fracb{\omega}{\omega_r}^{3/2}\hspace{-3mm} N^{1/2},
\end{align}
\item For fermions at zero temperature, using \eqref{ef} to set $\efht \gg  \hbar\omega_T$, we get
\begin{align}
    r_m \le \frac{1}{\sqrt{2}}\fracb{15\sqrt{\pi}N}{8}^{2/5} \frac{\ob}{\omega_r} \sigma_0
    \approx  1.14 \sigma_0 \frac{\ob}{\omega_r}N^{2/5}, 
\end{align}
\item In the high temperature limit we set $\kt \gg  \hbar\omega_T$ to get
\begin{align}
    r_m \le \frac{\sigma_0 }{\sqrt{2}}\frac{\kt}{\hbar\omega_r}\approx 0.707\sigma_0\frac{\kt}{\hbar\omega_r}.
\end{align}
\end{enumerate}
For the parameters of Figs.~\ref{f:loading} and \ref{f:loadingF}, the upper bounds for $r_m$ are $\sim 0.1\:{\rm mm}$ at $T=0$ and $\sim 1\:{\rm mm}$ at $T_c$ or $T_F$. By optimizing the theory we have developed here, much larger toroids would be possible.
\subsection{Investigating KZM using scale-invariance}
\begin{figure}[!t]
\includegraphics[width=3.4in]{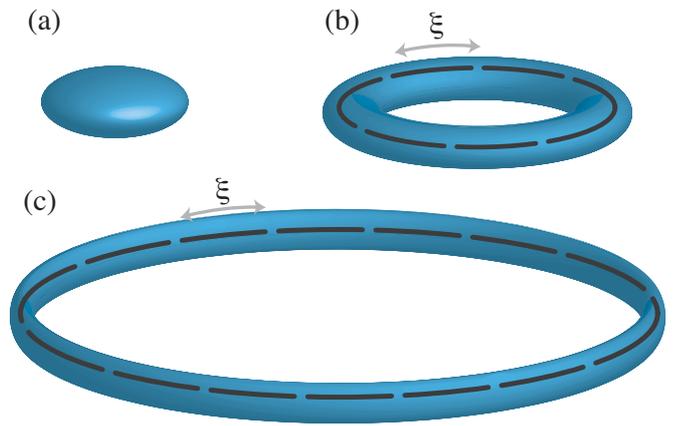}
 \caption{Equipotential surfaces in for the case of a (a) harmonic trap, (b) small toroidal trap, and (c) large toroidal trap. The correlation length is indicated (see text).\label{schematic1}}
\end{figure}

The following are desirable features of the scale invariant toroidal trap for studies of the Kibble-Zurek effect in an ultra-cold Bose gas:

\paragraph{System control:} The trap should reliably produce systems near the critical temperature and have good control over quenching the system through the critical point. Cooling in a harmonic trap and then loading using scale-invariance provides this feature, as a target degeneracy can be produced.
\paragraph{Suppression of post-quench dynamics:} An important practical issue in KZM experimental studies concerns the post-quench recombination dynamics of vortices: Evaporative cooling across the BEC transition in a harmonic trap is followed by rapid decay of the vortex density~\cite{BPAprivate}, associated with recombination of vortex anti-vortex pairs~\cite{Neely10a} and the decay of lone vortices~\cite{Rooney10a}. This can greatly complicate measurements of defect density scaling with quench rate. The same process in a toroidal trap will aggregate defects into a macroscopic persistent current~\cite{Damski10a}. It may be easier in practice to measure these energetically stabilized vortices than to characterize the decay of vortices in a harmonic trap.
\paragraph{Effective one dimensional quench:} A thermal gas quenched in a large toroid will not be quenched in the transverse degrees of freedom, because they respond much more rapidly than the azimuthal modes. Above $T_c$ the relaxation time for mode with energy $\epsilon$ is well estimated as \cite{Blakie08a}
\begin{equation}
\tau_j \sim\frac{k_B T}{|\mu-\epsilon_j|\gamma},
\end{equation}
where $\epsilon_0=0$ corresponds to the ground state, and $\hbar\omega_K\ll \hbar\omega_T,\hbar\omega_z$ will typically hold.  Here $\gamma$ is the damping rate arising from thermal cloud interactions. The system can be out of equilibrium azimuthally, but still in equilibrium with respect to transverse degrees of freedom with much higher energy. 
\paragraph{Correlation length:} An essential quantity of interest in KZM is the correlation length $\xi$, over which the first order correlation function decays. For an ideal homogeneous gas in thermal equilibrium at $T_c<T$, a semi-classical calculation gives \cite{Huang63}
\begin{equation}
 \xi^2=\frac{\hbar^2}{2m|\mu|}.
\end{equation}
 In a local density description of the trapped gas, we have $\xi(\bx)^2=\hbar^2/2m|\mu(\bx)|$, with $\mu(\bx)=\mu-V(\bx)$ the local chemical potential.
A scale invariant ideal gas will have a scale invariant chemical potential at the trap center, and hence the equilibrium correlation length will also be scale invariant there. Thus, defect formation for a range of tori with different perimeters could be explored for a given initial temperature and chemical potential of the gas (and therefore identical correlation length). For a larger toroid a larger number of correlation lengths could be frozen in during the quench (see Fig.~\ref{schematic1}). This could provide a means to make persistent currents with high winding number.
\subsection{Conclusions}
In this work we have established the scale invariance property of a harmonic Gaussian trap, extending the largely ideal Bose gas treatment of \cite{Bradley09a} to systems of interacting bosons and spin polarized fermions by analyzing the density of states. We have assessed the level of heating incurred through adiabatic loading of bosons with interactions, and of spin polarized fermions.
\par
Specifically, we have provided a formal proof of scale invariance of the general harmonic-Gaussian toroidal trap, demonstrating that large scaling transformations can be applied to the toroid without altering thermal equilibrium properties due to the invariant nature of trap geometry. Focusing on adiabatic loading from a harmonic trap, we have considered a range of possible tori determined by the Gaussian height $V_0$ and the energy $V_\sigma=m\omega_r^2\sigma_0^2/2$ which are the two energies defining the general trap. Analytical and numerical results for a single component Fermi gas, and for an interacting Bose gas are compared, allowing the relative temperature change under isentropic loading to be determined. 

\emph{Bose gas:} The Bose gas always exhibits a regime where degeneracy increases slightly, occurring for initial reduced temperatures lying in the range $0<t_H\lesssim 0.4$. We also observe that for $t_H\lesssim 1$ the degeneracy change that occurs using loading is quite small, which should allow loading of a well-characterized system in a harmonic trap into a toroid with the desired temperature. 

\emph{Fermi gas:}  Loading into a toroid always causes heating. For the special case of loading into a harmonic toroid, the smallest heating that may occur is 20\%, reflecting the fact that the angular position coordinate is not available for thermalization in the toroidal system. Also, the Fermi energy itself sets an important energy scale for the onset of scale invariance, and that the Fermi gas shows a slower approach to the scale invariant regime than an equivalent Bose gas, with respect to increasing $V_\sigma$.

We emphasize that, for finite $N$, the system is always scale-invariant as $V_0/V_\sigma\to\infty$, since the energy scale determining invariance, $V_0-V_m$, tends to $\infty$.
This property suggests that the harmonic Gaussian trap may provide a means to generate degenerate systems characterized by a macroscopic toroidal perimeter, or to make persistent currents of high winding number.

\begin{acknowledgments}
We thank Wojciech Zurek, Matthew Davis, Brian Anderson and Boris Baeumer for stimulating discussions. This work is supported by the New Zealand Foundation for Research, Science and Technology contracts NERF-UOOX0703 and UOOX0801, Marsden contract 09-UOO-093 and the New Zealand Tertiary Education Commission TADS.
\end{acknowledgments}
 
\appendix
\section{Density of states series expansion}\label{AppDOS}
\begin{figure}[!t] 
\includegraphics[width=3.4in]{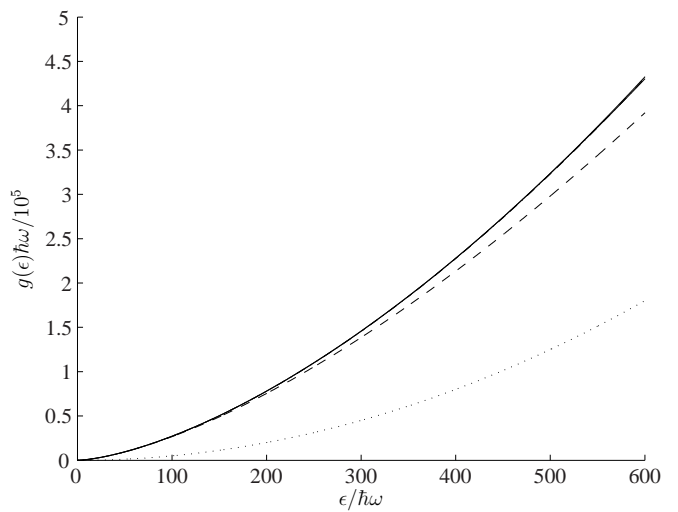}
\caption{\label{fulldos} Density of states for the parameters of Fig.~\ref{f:loading}(b) and with $V_0 = 5V_\sigma$: harmonic trap (dotted), series expansion with one (dashed), and two (solid) terms; with two terms, the series expansion and the exact numerical result (also solid) are indistinguishable on this scale.}
\end{figure}

Here we obtain a formal series expansion of the density of states in the toroidal potential, in the scale-invariant regime. We make use of a decomposition of the general density of states problem into a sequence of convolutions of densities of states corresponding to each additive term in the single particle energy \cite{Baillie2009b}. We can write, for example 
\begin{eqnarray}
    g(\epsilon) &= \int_0^{\epsilon}g_K(K)\gt(\epsilon-K) d K, \label{e:dosconv}
\end{eqnarray} 
and work with the \emph{trap density of states} $g_{\rm trap}(V)$ separately from the \emph{kinetic density of states} $g_K(K)$. We can further decompose the trap density of states into radial and axial parts 
\begin{eqnarray}
    \gt(V) &=& 2\pi\int_0^V g_r(V_r)g_{z}(V-V_r)d V_r.\label{e:doszrconv} 
\end{eqnarray}
For $V_0 > V_\sigma$, $V_m = V_\sigma\left[1+\ln(V_0/V_\sigma)\right]$ and the effective trap is
\begin{eqnarray}
 \ve(r) &=& \frac{m}{2} \omega_r^2 r^2 + V_0 e^{-r^2/\sigma_0^2} - V_m.\label{e:vexp}
\end{eqnarray}
Our main task is to find a useful expression for $g_r(V)$
\begin{align}
  g_r(V) &= \pdiff{}{V} \int_{\ve(r)\le V} r d r = \half \pdiff{}{V} (R_+^2 - R_-^2)\label{e:grVdiff} \\
 &= \frac{1}{m\omega_r^2}\left[\frac{1}{1 - \frac{V_0}{V_\sigma} e^{-R^2_+ /\sigma_0^2}} - \frac{1}{1 - \frac{V_0}{V_\sigma} e^{-R^2_- /\sigma_0^2}}\right],\label{e:grV}
\end{align}
where $R_\pm$ are defined to satisfy $\ve(R_\pm) = V$, which we implicitly differentiate using \eqref{e:vexp}. This requires that  $0<R_-<R_+$, i.e.~$V<V_0-V_m$, which is a necessary condition for the following. We proceed by expanding the potential about the minimum
\begin{eqnarray}
V_{\rm eff}(r)  &=& \sum_{j=2}^\infty c_j(r-r_m)^j,\label{e:vseries}
\end{eqnarray}
where, for example, $c_2 = m\omega_T^2/2$ and $c_3 = \omega_T(3\omega_r^2-\omega_T^2)\sqrt{m^3/V_\sigma}/6$.  
Inverting the series \cite{Chernoff47a}, we find, to $\od{V/V_\sigma}^2$
\begin{align}
  R_\pm - r_m &= \sqrtf{2V_\sigma}{m\omega_T^2} \Bigg\{\pm \sqrtf{V}{V_\sigma} 
 + \frac{1}{3\sqrt{2}}\Bigg[1  - 3\fracb{\omega_r}{\omega_T}^2  \Bigg] \frac{V}{V_\sigma}\notag\\
 &\pm \frac{1}{18}\Bigg[1-6\fracb{\omega_r}{\omega_T}^2+18\fracb{\omega_r}{\omega_T}^4 \Bigg]\fracb{V}{V_\sigma}^{3/2}  \Bigg\}.\label{e:Rseries}
\end{align}
Using Eq.~(\ref{e:grV}) we find, to $\od{V/V_\sigma}^{5/2}$
\begin{align}
  g_r(V) &= \frac{\sqrt{2}}{m\omega_r^2}\left[
  \sqrtf{V_\sigma}{V} 
  + \frac{1}{6}  \sqrtf{V}{V_\sigma} 
  + \frac{1}{216}\fracb{V}{V_\sigma}^{3/2} \right],
\end{align}
and \eqref{e:doszrconv} gives trap density of states, to $\od{V/V_\sigma}^3$ 
\begin{align}
  \gt(V) &= \frac{4\pi^2\sqrt{V_\sigma}}{(m\omega^2)^{3/2}}\left[1 +\frac{1}{12}\frac{ V}{V_\sigma} + \frac{1}{576}\fracb{ V}{V_\sigma}^2 \right],\label{e:gtvpoly}
\end{align}
using $g_z(V_z) = \sqrt{2/m\omega_z^2 V_z}$. Finally, we make use of $g_K(K) = m^{3/2}\sqrt{K/2}/\pi^2\hbar^3$ and \eqref{e:dosconv}
 to find the total density of states expansion \eqref{analyticDOS}. The expansion for the density of states is compared with  exact numerical results in Fig.~\ref{fulldos} .
\section{Proof of scale invariance\label{s:AppSI}}
In Appendix \ref{AppDOS}, we derived the first few terms of the series expansion of the density of states which were scale-invariant. Here we demonstrate the scale-invariant regime more generally. We define $R_\pm$ by $\ve(R_\pm) = \epsilon$ (with $R_- = 0$ for $\epsilon>V_0-V_m$), so that, for $V_0\ge V_\sigma$,
\begin{align}
  \epsilon = \frac{m}{2} \omega_r^2 R_\pm^2 + V_0 e^{-R_\pm^2/\sigma_0^2} 
  - \frac{m}{2}\omega_r^2\sigma_0^2\left[1+\ln(V_0/V_\sigma)\right].\label{e:Rdef} 
\end{align}
Implicitly differentiating, keeping $\epsilon$ and $\sigma_0$ constant, we find $\pdiffl{R_\pm^2}{V_0} = \sigma_0^2/V_0$ is not dependent on $\epsilon$, either directly or via $R_\pm$ (which is a function of $\epsilon$). Note that for $V_0 < V_\sigma$ (the dimple or flattened trap), $\pdiffl{R_\pm^2}{V_0}$ is a function of $R_\pm$ so that the following doesn't apply. For $\epsilon>V_0-V_m$, $R_-=0$, so  $\pdiffl{R_\pm^2}{V_0} = 0$ in that region.

From Eq. (37) of \cite{Bradley09a}, using \eqref{e:vexp}
\begin{align}
  g(\epsilon) &= \frac{1}{\hbar^3} \frac{m}{\omega_z}\int_{R_-}^{R_+} d r \, r \left[ \epsilon - \ve(r) \right]\\
  &= \frac{m}{2\hbar^3\omega_z}\left[r^2(\epsilon + V_m) - \frac{m}{4}\omega_r^2 r^4 +  V_0 \sigma_0^2e^{-r^2/\sigma_0^2} \right]_{R_-}^{R_+}. \label{e:combdos}
\end{align}
So,
\begin{align}
  \pdiff{g(\epsilon)}{V_0} &= \frac{m}{2\hbar^3\omega_z}\left[\pdiff{(r^2)}{V_0}\left\{
  \epsilon - \ve(r)\right\} +  r^2 \frac{V_\sigma}{V_0} + \sigma_0^2 e^{-r^2/\sigma_0^2} 
 \right]_{R_-}^{R_+}\nonumber\\
 &= \begin{cases}
   \frac{V_\sigma}{(\hbar\omega)^3V_0}\left(\epsilon+V_m - V_0\right) & {\rm if}\;R_-=0 \\
 0 & \text{otherwise}.
 \end{cases}\label{e:pdiffgev0}
\end{align}
Hence $g(\epsilon)$ does not vary with $V_0$ for $R_- > 0$, i.e. $\epsilon < V_0-V_m$ and the system is scale invariant for $\kt \ll  V_0 - V_m$  (note that we do \emph{not} require $\epsilon \ll V_0-V_m$).  Interestingly, if we define  $V^*_0$ to satisfy $V_0^* = \epsilon+V_m$, then we can integrate \eqref{e:pdiffgev0}
 to get the following exact correction to the scale invariant density of states, $\gsi(\epsilon)$ from \eqref{analyticDOS}, for $V_0 > V_\sigma$ and $\epsilon > V_0-V_m$
\begin{align}
    g(\epsilon) &= \gsi(\epsilon) + \frac{V_\sigma}{(\hbar\omega)^3}\left[ (\epsilon + V_\sigma)\ln \frac{V_0}{V_0^*}
 - V_0 + V^*_0 \bsplit + \half[V_\sigma] \left(\ln^2\frac{V_0}{V_\sigma} - \ln^2\frac{V^*_0}{V_\sigma}\right) \right].
 \end{align}
If we differentiate $g_r(V)$ [Eq.~\eqref{e:grVdiff}]
\begin{align}
  \pdiff{ g_r(V) }{V_0} &= \half \pdiff{}{V} \left(\pdiff{R_+^2}{V_0} - \pdiff{R_-^2}{V_0}\right) = 0, 
\end{align}
then use \eqref{e:doszrconv}, we find that the trap density of states, $\gt(V)$, is also scale invariant for $V<V_0-V_m$.

\end{document}